%% file: run.tex
\documentstyle[12pt,draft]{article}

 \newcounter{abceqn}

 \newcounter{abcfig}

\newcommand{\vth}{\vartheta}

\newcommand{\Ga}{\Gamma}

\newcommand{\pa}{\partial}

\newcommand{\tth}{\tilde{\theta}}
\newcommand{\tv}{\tilde{v}}

\newcommand{\tq}{\tilde{q}}

\newcommand{\e}{\epsilon}

\newcommand{\k}{\kappa}
\newcommand{\ga}{\gamma}

\newcommand{\dl}{\delta}

\newcommand{\th}{\theta}
\newcommand{\Th}{\Theta}
\newcommand{\ra}{\rightarrow}

\newcommand{\Sg}{\Sigma}

\newcommand{\z}{\zeta}

\newcommand{\La}{\Lambda}
\newcommand{\la}{\lambda}

\newcommand{\nid}{\noindent}

\def\maprightu#1{\smash{
    \mathop{\longrightarrow}\limits^{#1}}}
\def\maprightd#1{\smash{
    \mathop{\longrightarrow}\limits_{#1}}}
\def\mapdownl#1{
    \llap{$\vcenter{\hbox{$\scriptstyle#1$}}$}\Big\downarrow}
\def\mapdownr#1{\Big\downarrow
    \rlap{$\vcenter{\hbox{$\scriptstyle#1$}}$}}

\renewcommand{\theequation}{\thesection.\arabic{equation}}
\newcommand{\eqnsection}[1]{
	\section{#1}
	\setcounter{equation}{0}
	\renewcommand{\theequation}{\thesection.\arabic{equation}}
	\setcounter{figure}{0}
	\renewcommand{\thefigure}{\thesection.\arabic{figure}}
	\setcounter{remark}{0}
	\renewcommand{\theremark}{\thesection.\arabic{remark}}
	\setcounter{theorem}{0}
	\renewcommand{\thetheorem}{\thesection.\arabic{theorem}}
	\setcounter{lemma}{0}
	\renewcommand{\thelemma}{\thesection.\arabic{lemma}}
}

\title{{\bf Chaos and Shadowing Lemma for Autonomous Systems 
of Infinite Dimensions}}

\author{  \\ \\ \\ \\ 
Yanguang (Charles)\ \ Li  \thanks{This work is partially supported 
by the Guggenheim Fellowship. MSC: 35  46}
\\ \\ Department of Mathematics 
\\ University of Missouri \\ 
Columbia, MO 65211, USA \\ e-mail: cli@math.missouri.edu \\ 
phone: 573-884-0622 \\ fax: 573-882-1869}

\date{\today}

\renewcommand{\theequation}{\thesection.\arabic{equation}}

\begin{document}
\bibliographystyle{plain}
\maketitle
\newpage
\begin{abstract}    
For finite-dimensional maps and periodic systems, Palmer rigorously 
proved Smale horseshoe theorem using shadowing lemma in 1988 \cite{Pal88}.
For infinite-dimensional maps and periodic systems, such a proof was 
completed by Steinlein and Walther in 1990 \cite{SW90}, and Henry 
in 1994 \cite{Hen94}. For finite-dimensional autonomous systems, 
such a proof was accomplished by Palmer in 1996 \cite{Pal96}. 
For infinite-dimensional autonomous systems, the current article offers 
such a proof. First we prove an Inclination Lemma to set up a coordinate 
system around a pseudo-orbit. Then we utilize graph transform and 
the concept of persistence of invariant manifold, to prove the existence 
of a shadowing orbit.
\end{abstract}

\newtheorem{lemma}{Lemma}
\newtheorem{theorem}{Theorem}
\newtheorem{corollary}{Corollary}
\newtheorem{remark}{Remark}
\newtheorem{definition}{Definition}
\newtheorem{proposition}{Proposition}
\newtheorem{assumption}{Assumption}

\newpage
\tableofcontents

\newpage
\input shadow

\newpage
\bibliography{shadowing}

\end{document}

%% file: shadow.tex
\eqnsection{Introduction}

Since the invention of Shadowing Lemma by Anosov in 1967, it has been applied 
in a variety of situations. Starting from 1984 \cite{Pal84}, Palmer had been 
trying to use shadowing lemma to rigorously prove Smale horseshoe theorem. In 
1988, he successfully completed such a proof \cite{Pal88}. This proof works 
for finite-dimensional maps and periodic systems. Since then, such an 
application of shadowing lemma had been amplified in different situations 
including infinite dimensions, non-invertible maps etc. \cite{CLP89}
\cite{CLP89a} \cite{HL86} \cite{Bla86} \cite{SW89} \cite{Zen95} \cite{Zen97}.
For infinite-dimensional maps and periodic systems, rigorous proof of 
Smale horseshoe theorem using shadowing lemma was completed by Steinlein 
and Walther in 1990 \cite{SW90}, and Henry in 1994 \cite{Hen94}. Such a 
proof for autonomous systems had been elusive for a long time. For 
finite-dimensional autonomous systems, such a proof was completed by 
Palmer in 1996 \cite{Pal96}. Applications of such shadowing lemma 
for finite-dimensional autonomous systems had been amplified \cite{CKP95} 
\cite{CKP97}. Symbolic labeling of orbits for finite-dimensional 
autonomous systems had been investigated by Silnikov \cite{Sil67}. 
For infinite-dimensional autonomous systems, the current article offers 
a proof of Smale horseshoe theorem using shadowing lemma. We first set up 
a pseudo-orbit, then prove an Inclination Lemma to set up a proper 
coordinate system around the pseudo-orbit. Finally we use graph transform 
and the concept of persistence of invariant manifold of Fenichel \cite{Fen71},
to prove the existence of a shadowing orbit.

Our interest in such a proof of Smale horseshoe theorem for 
infinite-dimensional autonomous systems, lies in the development of the 
theory of chaos in partial differential equations \cite{LM94} \cite{LMSW96}
\cite{Li99} \cite{Li01a}. Indeed, we have presented as an example, a 
derivative nonlinear Schr\"odinger equation.

The article is organized as follows: Section 1 is the Introduction. 
Section 2 is the Setup of Assumptions. In Section 3, we set up the 
pseudo-orbits. In Section 4, we prove an Inclination Lemma, and set up 
a proper coordinate system around a pseudo-orbit. In Section 5, we prove a 
Shadowing Lemma. In Section 6, we prove Smale Horseshoe Theorem for a 
Poincar\'e return map. In Section 7, we present an example: A Derivative 
Nonlinear Schr\"odinger Equation. In Section 8, we present an example
of periodic systems: A Periodically Perturbed Sine-Gordon Equation.

\eqnsection{The Setup}

Let $B$ be a Banach space on which an autonomous flow is defined.
We set up the assumptions as follows.
\begin{itemize}
\item {\bf Assumption (A1):} There exist a hyperbolic limit cycle
$S$ and a transversal homoclinic orbit $\xi$ 
asymptotic to $S$. As curves, $S$ and $\xi$ are $C^{3}$.
\item {\bf Assumption (A2):} The Fenichel fiber theorem is valid at $S$. 
That is, there exist a family of unstable Fenichel fibers
$\{ {\cal F}^{u}(q): \  q \in S\}$ and a family of stable Fenichel 
fibers $\{ {\cal F}^{s}(q): \  q\in S\}$. For each fixed $q\in S$,
${\cal F}^{u}(q)$ and ${\cal F}^{s}(q)$ are $C^{3}$ submanifolds.
${\cal F}^{u}(q)$ and ${\cal F}^{s}(q)$ are $C^{2}$ in $q,
\forall q\in S$. The unions $\bigcup_{q\in S}{\cal F}^{u}(q)$ and
$\bigcup_{q\in S}{\cal F}^{s}(q)$ are the unstable and stable 
manifolds of $S$. Both families are invariant, i.e.
\[
F^{t}({\cal F}^{u}(q))\subset
{\cal F}^{u}(F^{t}(q)),  \forall\ t \leq 0, q\in S,
\]
\[
F^{t}({\cal F}^{s}(q))\subset {\cal F}^{s}(F^{t}(q)), 
\forall\ t \geq 0,  q \in S, 
\]
where $F^{t}$ is the evolution operator. There are positive constants
$\k$ and $\widehat{C}$ such that $\forall q\in S$, $\forall 
q^{-}\in {\cal F}^{u}(q)$ and $\forall q^{+}\in
{\cal F}^{s}(q)$,
\[
\| F^{t}(q^{-})-F^{t}(q)\| \leq
\widehat{C}e^{\k t}\| q^{-}-q\|,  \forall \ t \leq 0\ ,
\]
\[
\| F^{t}(q^{+})-F^{t}(q)\| \leq \widehat{C}e^{-\k t}\| q^{+}-q\|, 
\forall \ t \geq 0\ .
\]
\item {\bf Assumption (A3):} $F^{t}(q)$ is $C^{0}$ in $t$, for
$t\in (-\infty ,\infty)$, $q\in B$. For any fixed $t\in 
(-\infty ,\infty )$, $F^{t}(q)$ is a $C^{2}$ diffeomorphism on
$B$.
\end{itemize}

\begin{remark}
Notice that we do not assume that as functions of time, $S$ and $\xi$ 
are $C^3$ , and we only assume that as curves, $S$ and $\xi$ 
are $C^3$.
\end{remark}

\eqnsection{The Pseudo-Orbits \label{PO}}

The building blocks of the pseudo-orbits are what we call Loop-0 and 
Loop-1.
\begin{definition}
Loop-0, denoted by $\eta_{0}$, is defined to be the $m$-times circulation of
the limit cycle $S$, where $m$ is to be determined. See Figure \ref{lp0} 
for an illustration.
\end{definition}
To define Loop-1, choose points $p_{s}$, $p_{c}$, and $p_{u}$ on $S$, such 
that the arc-lengths of $\widehat{p_{s}p_{c}}$ and $\widehat{p_{c}p_{u}}$ are
equal to $\delta$, where $\dl$ is a small parameter to be determined. Let 
$\hat{p}_u$ be one of the points of
intersection $\xi \cap {\cal F}^{u}(p_{u})$ such that $\|
\hat{p}_{u}-p_{u}\| \sim {\cal O}(\dl^\nu) \ \mbox{as}\ \dl \ra 0$, and
$\hat{p}_{s}$ be one of the points of intersection $\xi \cap
{\cal F}^{s}(p_{s})$ such that $\| \hat{p}_{s}-p_{s}\| \sim 
{\cal O}(\dl^\nu) \ \mbox{as}\ \dl \ra 0$, where $\nu \geq 4$ ($\nu$ will 
be determined again later). Let $\z =\z_{u}\cup \z_{s}$ be a curve
connecting $\hat{p}_{s}$, $p_{c}$, and $\hat{p}_{u}$.
$\z_{s}$ lies in the stable manifold $W^{s}(S)$ of $S$, and
connects $\hat{p}_{s}$ and $p_{c}$. $\z_{u}$ lies in the unstable manifold 
$W^{u}(S)$ of $S$, and connects $p_{c}$ and $\hat{p}_{u}$. 
Let $\hat{\xi}$ be the connected arc-portion
of $\xi$, which links $\hat{p}_{u}$ and $\hat{p}_{s}$.
Then the union $\eta_{1}=\hat{\xi}\cup \z$ is a loop.
We choose $\z$ such that $\eta_1$ is $C^3$, $\widehat{p_sp_c}\cup \z_u$
is $C^3$, and $\z_s \cup \widehat{p_cp_u}$ is $C^3$. See Figure \ref{lp1} 
for an illustration.
\begin{figure}
\vspace{1.5in}
\caption{An Illustration of Loop-0.}
\label{lp0}
\end{figure}
\begin{figure}
\vspace{1.5in}
\caption{An Illustration of Loop-1.}
\label{lp1}
\end{figure}
\begin{definition}
Loop-1 is defined to be $\eta_{1}=\hat{\xi}\cup \z$.
\end{definition}
To define the pseudo-orbits, first we introduce sequences of symbols.
\begin{definition} 
Let $\Sigma $ be a set that consists of
elements of the doubly infinite sequence form:
\[
a=(\cdots a_{-2}a_{-1}a_{0}, a_{1}a_{2}\cdots ),
\]
where $a_{k}\in \{0,1\}$, $k\in Z$. We introduce a topology
in $\Sigma$ by taking as neighborhood basis of
\[
a^{*}=(\cdots a^{*}_{-2}a_{-1}^{*}
a^{*}_{0},a^{*}_{1}a^{*}_{2}\cdots ),
\]
the set
\[
B_{j}=\{a\in \Sigma \ |\ \ a_{k}=a^{*}_{k} \ (|k|<j)\}
\]
for $j=1,2,\ldots $. This makes $\Sigma$ a topological space. The
Bernoulli shift automorphism $\chi$ is defined on $\Sigma$ by
\[
\chi :\Sigma \mapsto \Sigma ,\quad \forall a\in \Sigma ,\ 
\chi (a)=b, \ \mbox{where}\ b_{k}=a_{k+1}.
\]
The Bernoulli shift automorphism $\chi$ 
exhibits sensitive  dependence on initial conditions,
which is a hallmark of chaos.
\end{definition}
To each $a_{k}\in \{ 0,1\}$, we associate the loop-$a_{k}$,
$\eta_{a_{k}}$. Then each doubly infinite sequence
\[
a=(\cdots a_{-2}a_{-1}a_{0},a_{1}a_{2}\cdots )
\]
is associated with a $\delta$-pseudo-obit
\[
\eta_{a}=(\cdots
\eta_{a_{-2}}\eta_{a_{-1}}\eta_{a_{0}},\eta_{a_{1}}\eta_{a_{2}}\cdots ).
\]
Since $\eta_a$ is a $C^3$ curve, we choose a parametrization of 
$\eta_a$: $\eta_a = \eta_a (\tau)$, $\tau \in R$, such that $\eta_a (\tau)$
is a $C^3$ function of $\tau$. 

\eqnsection{Coordinates along the Pseudo-Orbits and Inclination Lemma}

We define bundles along the limit cycle $S$ and the homoclinic orbit
$\xi$ as follows:
\begin{definition}
The bundles $E^u$, $E^c$, and $E^s$ along $S$ are defined by 
\[
E^u(q) ={\cal T}_q {\cal F}^u(q), \quad 
E^c(q) ={\cal T}_q S            , \quad 
E^s(q) ={\cal T}_q {\cal F}^s(q), \quad q \in S,
\]
where ${\cal T}_q$ indicates the tangent space at $q$.
\end{definition}
See Figure \ref{cl0} for an illustration. $E^u$ and $E^s$ provide a coordinate 
system for a tubular neighborhood of $S$, that is, any point in this 
neighborhood has a unique coordinate 
\[
(\tv^s,\tth,\tv^u), \quad \tv^s \in E^s(\tth), \quad \tv^u \in E^u(\tth),
\quad \tth \in S.
\]
Fenichel fibers provide another coordinate system for the tubular 
neighborhood of $S$. For any $\th \in S$, the Fenichel fibers ${\cal F}^s(\th)$
and ${\cal F}^u(\th)$ have the expressions
\[
\left \{ \begin{array}{l} \tv^s = v^s , \\ \tth= \th +\Th_s(v^s,\th), \\
\tv^u = V_s(v^s,\th), \end{array} \right.
\quad \quad \mbox{and}\quad \quad 
\left \{ \begin{array}{l} \tv^u = v^u , \\ \tth= \th +\Th_u(v^u,\th), \\
\tv^s = V_u(v^u,\th), \end{array} \right.
\]
where $v^s$ and $v^u$ are the parameters parametrizing ${\cal F}^s(\th)$
and ${\cal F}^u(\th)$,
\[
\Th_z(0,\th)= {\pa \over \pa v^z}\Th(0,\th) = V_z(0,\th)= 
{\pa \over \pa v^z}V_z(0,\th)= 0, \quad z=u,s;
\]
and $\Th_z(v^z,\th)$ and $V_z(v^z,\th)$ ($z=u,s$) are $C^3$ in $v^z$ 
and $C^2$ in $\th$. The coordinate transformation from ($v^s,\th,v^u$) 
to ($\tv^s, \tth,\tv^u$)
\[
\left \{ \begin{array}{l} \tv^s = v^s +V_u(v^u,\th), \\ 
\tth= \th +\Th_u(v^u,\th)+\Th_s(v^s,\th), \\
\tv^u =v^u+ V_s(v^s,\th), \end{array} \right.
\]
is a $C^2$ diffeomorphism. In terms of the Fenichel coordinate 
($v^s,\th,v^u$), the Fenichel fibers coincide with their tangent 
spaces. From now on, we always work with the Fenichel coordinate 
($v^s,\th,v^u$).
\begin{definition}
Let $q^u,q^s \in \xi$, and $q \in S$, such that $q^z \in {\cal F}^z(q)$,
$z=u,s$. The bundles $E^u$, $E^c$, and $E^s$ along $\xi$ are defined 
by 
\begin{eqnarray*}
& & E^u(F^t(q^u)) = DF^t(q^u)({\cal T}_{q^u} {\cal F}^u(q)), \quad 
t \in (-\infty,\infty), \\
& & E^c(p)={\cal T}_p\xi, \quad  p\in \xi, \\
& & E^s(F^t(q^s)) = DF^t(q^s)({\cal T}_{q^s} {\cal F}^s(q)), \quad 
t \in (-\infty,\infty). 
\end{eqnarray*}
\end{definition}
See Figure \ref{cl1} for an illustration. 
\begin{figure}
\vspace{1.5in}
\caption{An illustration of the bundles $E^u$, $E^c$, and $E^s$ along 
the limit cycle $S$.}
\label{cl0}
\end{figure}
\begin{figure}
\vspace{1.5in}
\caption{An illustration of the bundles along 
the homoclinic orbit $\xi$.}
\label{cl1}
\end{figure}
We have the following inclination lemma. 
\begin{lemma}[Inclination Lemma]
For $\dl$ small enough, there exists a $\nu_0>0$ ($\nu_0$ depends 
upon $\dl$), such that for any $q \in S$, let $q^u \in \xi \bigcap 
{\cal F}^u(q)$ and $q^s \in \xi \bigcap {\cal F}^s(q)$,
\begin{enumerate}
\item When $\| q^s -q\| \sim {\cal O}(\dl^\nu)$, $\nu \geq \nu_0$, 
$E^u(q^s)\oplus E^c(q^s)$ is $\dl^3$-close to $E^u(q)\oplus E^c(q)$.
\item When $\| q^u -q\| \sim {\cal O}(\dl^\nu)$, $\nu \geq \nu_0$, 
$E^s(q^u)\oplus E^c(q^u)$ is $\dl^3$-close to $E^s(q)\oplus E^c(q)$.
\end{enumerate}
\label{Inc}
\end{lemma}

Proof: Let $q_1^s \in \xi \bigcap {\cal F}^s(q)$ such that 
$\| q_1^s -q\| \sim {\cal O}(\dl^4)$. Notice that ${\cal F}^s(q) =
E^s(q)$. Let $v_1 \in E^u(q_1^s)\oplus E^c(q_1^s)$, $\| v_1 \| =1$. 
We represent $v_1$ in the frame ($E^s(q), E^u(q)\oplus E^c(q)$), 
\[
v_1 =(v_1^s, v_1^{uc}),
\]
where $\| v_1^{uc}\| \neq 0$, since $\xi$ is a transversal homoclinic 
orbit. Let $\la_1 = \| v_1^{s}\|\bigg / \| v_1^{uc}\|$. The transversality 
of $\xi$ implies that $\la_1$ has an upper bound for all $v_1$ \cite{Hen94}.
Since $S$ is compact, $\la_1$ has an upper bound for all $q\in S$. Let
$q_n^s$ be the consecutive intersection points between $\xi$ and 
${\cal F}^s(q) = E^s(q)$, $q_n^s = F^{T_*}(q_{n-1}^s)$, where $T_*$ 
is the period of the limit cycle $S$. Let
\begin{eqnarray*}
(\hat{v}_2^s, \hat{v}_2^{uc}) &=& DF^{4mT_*}(q_1^s)v_1 =
DF^{4mT_*}(q_1^s)(v_1^s, 0) + DF^{4mT_*}(q)(0,v_1^{uc}) \\
& & + \bigg [ DF^{4mT_*}(q_1^s)-DF^{4mT_*}(q) \bigg ](0,v_1^{uc}).
\end{eqnarray*}
Also let 
\[
(r^s, r^{uc}) = \bigg [ DF^{4mT_*}(q_1^s)-DF^{4mT_*}(q) \bigg ](0,v_1^{uc}).
\]
Then 
\[
\hat{v}_2^s = DF^{4mT_*}(q_1^s)v_1^s + r^s, \quad 
\hat{v}_2^{uc}= DF^{4mT_*}(q)v_1^{uc} +r^{uc}.
\]
We choose $m$ large enough such that the constant $\widehat{C}$ in 
Assumption (A2) satisfies $\widehat{C} \leq e^{\k m T_*}$, and 
\[
DF^{4mT_*}(q)v_1^{uc} \geq 2 e^{-\k m T_*} \|v_1^{uc}\|.
\]
For such fixed $m$, choosing $\dl$ small enough, we have 
\[
\| DF^{4mT_*}(q_1^s)-DF^{4mT_*}(q) \| \leq \dl^3 e^{-\k m T_*}.
\]
Then
\begin{eqnarray*}
& & \| \hat{v}_2^s \| \leq e^{-3\k m T_*}\| v_1^s\|+ \dl^3 e^{-\k m T_*} 
\|v_1^{uc}\|, \\
& & \| \hat{v}_2^{uc} \| \geq 2e^{-\k m T_*}\|v_1^{uc}\|- \dl^3 e^{-\k m T_*} 
\|v_1^{uc}\| \geq e^{-\k m T_*}\|v_1^{uc}\|.
\end{eqnarray*}
Thus
\[
\la_2 = \| \hat{v}_2^s \| \bigg / \| \hat{v}_2^{uc} \|
\leq e^{-2\k m T_*} \la_1 + \dl^3.
\]
Iterating the argument, we obtain
\[
\la_N \leq e^{-2\k (N-1)m T_*} \la_1 + \dl^3 \sum_{l=0}^{N-2} e^{-2\k lm T_*}.
\]
For such fixed $\dl$, when $N$ is large enough,
\[
\la_N \leq 4 \dl^3.
\]
There exists $\nu_0>0$ such that
\[
\| F^{4(N-1)mT_*}(q^s_1)-q\| \sim {\cal O}(\dl^\nu), \quad \nu \geq \nu_0.
\]
Similarly for the case of $q^u$. The proof is completed. $\Box$
\begin{remark}
Inclination lemmas have been utilized in proving many significant theorems 
\cite{PM82} \cite{Wal87}. Here we show how to use inclination lemma to 
prove shadowing lemma in the autonomous case.
\end{remark}

Based up the fact in Lemma \ref{Inc}, $C^2$ bundles can be constructed 
along the pseudo-orbit $\eta_a$ as follows (cf: Figure \ref{lp1}): 
Since $E^u(\hat{p}_s) \oplus E^c(\hat{p}_s)$ is $\dl^3$-close to 
$E^u(p_s) \oplus E^c(p_s)$, we can construct $C^2$ bundle 
$E^u(p) \oplus E^c(p)$, $p \in \widehat{\hat{p}_sp_c}$ along the 
curve $\widehat{\hat{p}_sp_c}$ such that $E^u(p) \oplus E^c(p)$ is 
$C^2$ along both the curve $p \in \widehat{\hat{p}_sp_c\hat{p}_u}$
and the curve $p \in \widehat{\hat{p}_sp_cp_u}$. $E^c(p)$ ($p \in 
\widehat{\hat{p}_sp_c}$) is the tangent space 
${\cal T}_p\widehat{\hat{p}_sp_c}$. For any $p \in \widehat{\hat{p}_sp_c}$,
$p \in {\cal F}^s(q)$, $q \in S$, we define 
$E^s(p) = {\cal T}_p{\cal F}^s(q)$. Similarly, we can define such 
bundles along the curve $\widehat{p_c\hat{p}_u}$. Thus we obtain 
$C^2$ bundles $E^u(p) \oplus E^c(p)$ and $E^s(p) \oplus E^c(p)$ along
$\eta_a$. We also need to rectify $E^u(p)$ and $E^s(p)$ along Loop-1. 
For any two subspaces $E_1$ and $E_2$, one can define the angle 
$\vth(E_1,E_2)$ as follows \cite{Hen94}:
\[
\vth(E_1,E_2) = \inf_{v_1,v_2} \bigg \{ \| v_1 -v_2 \| \ \bigg | \ 
v_1 \in E_1, v_2 \in E_2, \| v_1\|=\| v_2\|= 1\bigg \} .
\]
Since $E^u(p)$ ($p\in S$) is a transversal bundle along $S$, and $S$ is 
compact, $\vth(E^u(p),E^c(p))  \geq \hat{\vth} > 0 $ for all $p\in S$. 
Let $p_n^s$ be the consecutive intersection points of $\xi$ with 
${\cal F}^s(p)$. Then $E^c(p_n^s) \ra E^c(p)$ as $n \ra \infty$. Thus, 
there exists $n_1$, $\vth(E^u(p), E^c(p_n^s))\geq \frac{1}{2} \hat{\vth}$ 
for all $n \geq n_1$, and all $p \in S$, by the compactness of $S$. 
Let $p^+\in \xi$, such that when $\dl$ is sufficiently small, 
\[
\vth(E^u(p),E^c(\tilde{p})) \geq \frac{1}{3} \hat{\vth}, \quad
\mbox{for all}\ \tilde{p} \in \widehat{p^+p_c}, \quad 
\tilde{p} \in {\cal F}^s(p), \quad p \in S.
\]
See Figure \ref{rect} for an illustration.
\begin{figure}
\vspace{1.5in}
\caption{A rectification of the transversal bundle along Loop-1.}
\label{rect}
\end{figure}
We can choose $p^+$ close enough to $p_0 \in S$, $p^+ \in 
{\cal F}^s(p_0)$, such that $E^u(q^+) \oplus E^c(q^+)$ is 
${\cal O}(\frac{1}{300} \hat{\vth})$ close to $E^u(q_0) \oplus E^c(q_0)$,
$q^+ \in \widehat{p^+p_c}$, $q^+ \in {\cal F}^s(q_0)$, $q_0 \in S$. 
For any $\tv^u \in E^u(q_0)$, $\| \tv^u \| = 1$, $\tv^u$ has the 
representation in the frame ($E^s(q^+), E^u(q^+)\oplus  E^c(q^+)$)
\[
\tv^u = v^s + v^{uc} .
\]
Alternatively, $v^{uc}$ has the representation
\[
v^{uc} = -v^s + \tv^u
\]
in the frame ($E^s(q_0), E^u(q_0)\oplus  E^c(q_0)$), since 
$E^s(q_0)= E^s(q^+)$. Thus
\[
\| v^s \| \leq \frac{1}{300} \hat{\vth} \| \tv^u\| = \frac{1}{300} \hat{\vth},
\]
and
\[
1-\frac{1}{300} \hat{\vth} \leq \| v^{uc}\| \leq 1+ \frac{1}{300} \hat{\vth}.
\]
Let $v^c \in E^c(q^+)$, $\| v^c\| =1$, we have
\begin{eqnarray*}
\frac{1}{3} \hat{\vth} &\leq & \| \tv^u - v^c\| = \|  v^s + v^{uc} -
v^{uc} / \| v^{uc} \| + v^{uc} / \| v^{uc} \| - v^c \| \\
& & \leq \|  v^s \| + \bigg | 1- \frac{1}{\| v^{uc} \|} \bigg |
\| v^{uc} \| + \bigg \| v^{uc} / \| v^{uc} \| - v^c \bigg \|  \\
& & \leq \frac{1}{300}\hat{\vth} + \frac{1}{300}\hat{\vth} + 
\bigg \| v^{uc} / \| v^{uc} \| - v^c \bigg \|.
\end{eqnarray*}
Thus we have
\[
\bigg \| v^{uc} / \| v^{uc} \| - v^c \bigg \| \geq \frac{98}{300}\hat{\vth}.
\]
In fact, $v^{uc}$ is the projection of $\tv^u$ onto $E^u(q^+)\oplus  E^c(q^+)$.
All such $v^{uc}$'s span the projection $\tilde{E}^u(q^+)$ of $E^u(q_0)$ 
onto $E^u(q^+)\oplus  E^c(q^+)$. Thus
\[
\vth(\tilde{E}^u(q^+), E^c(q^+)) \geq \frac{98}{300}\hat{\vth}, \quad
q^+ \in \widehat{p^+ p_c}.
\]
At $q^+ \in \widehat{p^+ p_c}$, we replace $E^u(q^+)$ by $\tilde{E}^u(q^+)$,
and we use the same notation $E^u(q^+)$. $E^u(q^+)$ is still $C^2$ along 
both $\widehat{p^+ p_c\hat{p}_u}$ and $\widehat{p^+ p_cp_u}$. Similarly 
we can choose $p^- \in \xi$, construct $\tilde{E}^s(q^-)$ ($q^- 
\in \widehat{p_cp^-}$), replace $E^s(q^-)$ by $\tilde{E}^s(q^-)$, and 
use the same notation $E^s(q^-)$. Let $T_0 >0$ be the time such that 
$p^+ = F^{T_0}(p^-)$. Let $\tilde{\xi}$ be the portion of $\xi$, 
$\tilde{\xi} = \bigcup_{0\leq t \leq T_0}F^t(p^-)$. Inside 
$E^u(q)\oplus  E^c(q)$ ($q \in \tilde{\xi}$), we choose a $C^2$ transversal 
bundle $\tilde{E}^u(q)$ along $\tilde{\xi}$, replace $E^u(q)$ by 
$\tilde{E}^u(q)$, and use the same notation $E^u(q)$. Similarly, inside 
$E^s(q)\oplus  E^c(q)$ ($q \in \tilde{\xi}$), we choose a $C^2$ transversal 
bundle $\tilde{E}^s(q)$ along $\tilde{\xi}$, replace $E^s(q)$ by 
$\tilde{E}^s(q)$, and use the same notation $E^s(q)$. This way, we construct
$C^2$ transversal bundles $E^u(q)$ and $E^s(q)$ along $\eta_a$.

\eqnsection{Shadowing Lemma}

We will use graph transform and the concept of persistence of 
invariant manifold, to establish a shadowing orbit \cite{Fen71}. 
In \cite{Fen71}, the estimates in the current case are not 
given in details. We will remedy that below.

Denote by $\widehat{E}$ the transversal bundle
\[
\widehat{E}=\{ (q, E^u(q), E^s(q))\ | \ q \in \eta_a \} ,
\]
which serves as a coordinate system around $\eta_a$. Using the parametrization 
$\eta_a = \eta_a(\tau),\ \tau \in R$, we can introduce the coordinate in 
a neighborhood of $\eta_a$:
\[
(\tau, x^u, x^s), \quad \mbox{where}\ \tau \in R, x^u \in E^u(\eta_a(\tau)), 
x^s \in E^s(\eta_a(\tau)).
\]
In this coordinate system, the evolution operator $F^T$ has the representation:
\[
F^T(\tau, x^u, x^s)=(f(\tau, x^u, x^s), g^u(\tau, x^u, x^s), 
g^s(\tau, x^u, x^s)),
\]
where $T>0$ is a large time. First we define Lipschitz sections 
over $\eta_a$.
\begin{definition} Let $\Gamma_\epsilon$ be the space of sections of 
$\widehat{E}$:
\[
\Gamma_\epsilon = \{ \sigma \ | \ \sigma (\tau ) =(\tau 
,x ^u(\tau ), x ^s(\tau )), \tau \in R, \| x ^u(\tau ) \|
\leq \epsilon ,\| x^s(\tau) \| \leq \epsilon \}.
\]
We define the $C^0$ norm of $\sigma \in \Gamma_\epsilon$ as
\[
\| \sigma \|_{C^0}=\max \{ \sup_{\tau \in R}\| 
x^u(\tau )\| , \sup_{\tau
\in R}\| x^s(\tau )\| \}.
\]
\end{definition}
Then we define a Lipschitz semi-norm on $\Gamma_\epsilon$:
\[
\mbox{Lip}\ \{\sigma \} = \max \left\{ 
\sup_{|\tau _1-\tau _2|\leq \Delta} \frac{\| 
x^u(\tau_1)-x^u(\tau
_2)\| }{\|
\tau_1-\tau_2\|}\right.\left. ,\sup_{|\tau_1-\tau_2|\leq 
\Delta}\frac{\| x^s(\tau _1)-x^s(\tau
_2)\|}{\|\tau_1-\tau_2\|}\right \}
\]
for some small fixed $\Delta >0$. Let $\Gamma_{\epsilon, \ga }$ be a subset of $\Ga_\e$, 
\[
\Gamma_{\epsilon, \ga }=\{ \sigma \in \Gamma_\epsilon 
\ |\ \ \ \mbox{Lip}\ \{ \sigma \} \leq \ga \}.
\]

\begin{lemma}\label{compl} $\Gamma_{\epsilon , \ga }$ is closed under the 
$C^0$ norm.\end{lemma}

Proof: Assume that $\{ \sigma_j\}_{j=1,2,\ldots }$ is a Cauchy 
sequence in $\Gamma_{\epsilon , \ga }$ under the $C^0$ norm. Then
$\forall \tau \in R$,
$x^z_j(\tau ) (z=u,s)$ is a Cauchy sequence, which has a limit 
$x^z(\tau )$. Define a new section $\sigma $ by
\[
\sigma (\tau ) =(\tau ,x^u(\tau),x^s(\tau )).
\]
First we want to show that $\sigma \in
\Gamma_\epsilon$.  $\forall \tau \in R$,
$\forall j=1,2,\ldots $,
\[
\| x^z_j(\tau )\| \leq \epsilon ,\quad (z=u,s).
\]
Then $\forall \tau \in R$,
\[
\| x^z(\tau )\| \leq \epsilon, \quad (z=u,s).
\]
Thus $\sigma \in \Gamma_\epsilon$. Next, we 
want to
show that $\sigma \in
\Gamma_{\epsilon , \ga }.$ We know that for $\tau_1$ and $\tau_2$ such that
$|\tau_1-\tau _2|\leq \Delta$, and any $j=1,2,\ldots$,
\[
\frac{\|x^z_j(\tau _1)-x^z_j(\tau_2)\|}{\| 
\tau_1-\tau_2\|} \leq \ga ,\quad (z=u,s).
\]
Then
\[
\frac{x^z(\tau_1)-x^z(\tau_2)\|}{\|\tau_1-\tau_2\|} 
=\lim_{j\to \infty}
\frac{\| x^z_j(\tau_1)-x^z_j(\tau_2)\|}{\| \tau_1-\tau_2\|} 
\leq \ga ,\quad (z=u,s).
\]
Thus
\[
\mbox{Lip}\ \{\sigma \}=\max\left\{
\sup_{|\tau_1-\tau_2|\leq \Delta} \frac{\| x^u(\tau_1)-x^u(\tau_2)\|}{\|
\tau_1-\tau_2\|}\right. \left. ,\sup_{|\tau_1-\tau_2|\leq 
\Delta}\frac{\| x^s(\tau_1)-x^s(\tau_2)\|}{\|\tau_1-\tau_2\|}\right\} 
\leq \ga .
\]
The proof is completed. $\Box$

For any $\sigma \in \Gamma_{\epsilon , \ga }$,
\[
\sigma (\tau)=(\tau 
,x^u(\tau),x^s(\tau)),\quad \tau \in R,
\]
let $T>0$ be a large time, we define
the {\em{graph  transform}} $G$ as follows:
\begin{equation}
(G\sigma )(\tau )=(\tau ,x^u_1 (\tau 
),x^s_1(\tau )),
\label{gt} 
\end{equation}
where
\begin{eqnarray*}
& & f(\tau^-,x^u(\tau^-),x^s(\tau^-))=\tau ,\\
& & g^s(\tau^-,x^u(\tau^-),x^s(\tau^-))=x^s_1(\tau),\\ 
& & f(\tau ,x^u_1(\tau),x^s(\tau))=\tau^+\\ 
& & g^u(\tau,x^u_1(\tau),x^s(\tau))=x^u(\tau^+).
\end{eqnarray*}
See Figure \ref{gff} for an illustration.

First we shall prove the existence of a fixed point of $G$ in 
$\Gamma_{\epsilon , \ga }$. Then we will show that the graph of the fixed
point is an orbit. Thereby, we establish the existence of an orbit 
that $\epsilon$-shadows the pseudo-orbit. The following preliminary
lemmas are quoted from \cite{Die60}(pp.155 and pp.186).
\begin{figure}
\vspace{1.5in}
\caption{An Illustration of the Graph Transform.}
\label{gff}
\end{figure}
\begin{lemma}[Mean Value Theorem] Let $E_1$ and $E_2$ be two Banach 
spaces, $F$ a continuous mapping from a neighborhood of a segment
$\ell $ joining two points
$q_0,q_0+q_1$ of $E_1$, into $E_2$. If $F$ is differentiable at every 
point of $\ell$, then
\begin{equation}\label{MVI} \| F(q_0+q_1)-F(q_0)\| \leq
\|q_1\|\sup_{0\leq \alpha \leq 1}\| DF(q_0+\alpha 
q_1)\|.\end{equation}
\end{lemma}
\begin{lemma}[Taylor's Formula] Let $E_1$ and $E_2$ be two Banach spaces,
$\Omega$ an open subset of $E_1$, $F$ a $n$-times continuously 
differentiable mapping of $\Omega$ into $E_2$. Then, if the segment 
joining
$q_0$ and $q_0+q_1$ is in
$\Omega$, we have
\begin{eqnarray}
F(q_0+q_1) &=& F(q_0)+DF(q_0)\circ q_1+\frac{1}{2!}
D^2F(q_0)\circ q^{(2)}_1+\cdots +
\frac{1}{(n-1)!} D^{n-1}F(q_0)\circ q_1^{(n-1)} \nonumber \\
& & \quad +\left( \int^1_0\frac{(1-\alpha)^{n-1}}{(n-1)!}D^n 
F(q_0+\alpha q_1)d\alpha \right)\circ q_1^{(n)},
\label{texp}
\end{eqnarray}
where $q^{(k)}_1$ stands for $(q_1,\ldots ,q_1)$ ($k$-times).
\end{lemma}

Now we set up a tubular neighborhood of the closure $\xi \cup S$ of 
the homoclinic orbit $\xi$. For any $T$ such that $0<T<\infty$, $F^T$
is a $C^2$ diffeomorphism. $\forall q\in \xi \cup S$, choose $r>0$ such that
\begin{equation}\label{tube1} \| D^\ell F^{\pm T}(q_1)-D^\ell 
F^{\pm T}(q)\| <1,\quad (\ell = 1,2)\end{equation} for any $q_1\in
{\cal B}_q(r_q)$, the ball centered at $q$ with radius
$r_q$, i.e.
\[
{\cal B}_q(r_q)=\{ q_1\in B \ |\ \ \| q_1-q\| 
<r_q\}.
\]
Then $\bigcup_{q\in \xi \cup
S}{\cal B}_q(r_q)$ is an open covering of $\xi
\cup S$. Since $\xi \cup S$ is compact, there is a finite subcovering
$\bigcup_{q_j\in \xi \cup S(1\leq j\leq n)}{\cal B}_{q_j}(r_{q_j})$. 
For simplicity, we denote ${\cal B}_{q_j}(r_{q_j})$ and
$r_{q_j}$ by
${\cal B}_j$ and $r_j$ respectively $(1\leq j\leq N)$. Denote by
${\cal B}$ the collection
\begin{equation}\label{tube2} {\cal B}=\{ {\cal B}_j,(1\leq 
j\leq N)\}\end{equation} which is referred as the tubular neighborhood
of $\xi \cup S$. See Figure~\ref{tufig} for an illustration. $\forall 
q\in \xi \cup S$, we define
\[
d_q =\ \mbox{dist}\ (q,\partial 
{\cal B})=\max_{j_\ell}\{ \ \mbox{dist}\  (q,\partial
{\cal B}_{j_\ell}),\ q\in {\cal B}_{j_\ell}\}.
\]
Then we define
\begin{equation}\label{tube3} d=\ \mbox{dist}\  (\xi \cup S,\partial
{\cal B})=\inf_{q\in \xi \cup S}\{  \ \mbox{dist}\ (q,\partial
{\cal B})\}.\end{equation}
\begin{figure}
\vspace{1.5in}
\caption{An illustration of the tubular neighborhood ${\cal B}$.}
\label{tufig}
\end{figure}

\begin{lemma} $d>0$.\end{lemma}

Proof: $\forall q\in {\cal B}_j\cap (\xi \cup S)$, $(1\leq 
j\leq N)$, let $B^j_q$ be a ball centered at $q$ with radius equal
to $\frac{1}{2}\ \mbox{dist}\  (q,\partial {\cal B}_j)$. Then $\bigcup_{1\leq 
j\leq N}\bigcup_{q\in
{\cal B}_j\cap (\xi \cup S)}{\cal B}^j_q$ is an open covering of $\xi
\cup S$; thus there is a finite subcovering
$\bigcup_{k=1}^K\bigcup_{\ell=1}^{L_k}B^{j_k}_{q_\ell}$ for some 
positive integers
$K$ and $L_k$. Let $r_*>0$ be the smallest radius of the balls
$B^{j_k}_{q_\ell}$. Then $\forall q\in \xi \cup S$, $q\in 
B^{j_k}_{q_\ell}$ for some $k$ and $\ell$. Then $ \ \mbox{dist}\ (q,\partial
{\cal B})\geq r_*$. Thus $d\geq r_*>0$. The proof of the lemma is completed.
$\Box$

\nid
Let
\begin{equation}
\Lambda_\ell = \max_{+,-} \ \sup_{q\in
{\cal B}} \| D^\ell F^{\pm T}(q)\| 
= \max_{+,-}\ \max_{1\leq 
j\leq N}\ \sup_{q\in {\cal B}_j}\| D^\ell F^{\pm T}(q)\|,\quad
(\ell =1,2).\label{tube4}
\end{equation}
Then (\ref{tube1}) and (\ref{tube2}) imply that 
$\Lambda_\ell <\infty \ (\ell =1,2)$.
\begin{lemma}\label{Lian} $\forall \mu >0$, fix a $T$ large enough, 
and fix a $\epsilon$ small enough, if $\delta$ is small enough, then
\begin{eqnarray*}
& & (\La_1)^k \Pi_3^s <\frac{1}{2},\ \ (0\leq k\leq 2), \quad 
\Pi_\ell^s < \mu ,\ \ (\ell =1,2), \\ 
& & (\La_1)^k \widehat{\Pi}_2^u  <\frac{1}{2},\ \ (0\leq k\leq 2),
 \quad \Pi_\ell^u <\mu , \ \  (\ell =1,3),
\end{eqnarray*}
where $\| x^u\| \leq \epsilon$, $\| x^s\|\leq 
\epsilon $, $D_1=D_\tau$, $D_2=D_{x^u}$,
$D_3=D_{x^s}$, and 
\begin{eqnarray*}
& & \Pi ^s_\ell =\sup_{\tau ,x^u,x^s}\| D_\ell g^s(\tau ,x^u,x^s)\|, \ \ 
(\ell =1,2,3),\\ 
& & \Pi^u_\ell =\sup_{\tau ,x^u ,x^s}\| D_\ell 
g^u(\tau ,x^u,x^s)\| , \ \ (\ell =1,2,3),\\
& & \widehat{\Pi}^u_2=\sup_{\tau ,x^u,x^s}\| \{D_2g^u(\tau ,x^u,x^s)\}^{-1}\|.
\end{eqnarray*}
\end{lemma}

Proof: Since $E^s(q)$ and $E^u(q)$ ($q\in \eta_a$) are transversal 
bundles, the two inequalities involving $\Pi_3^s$ and $\widehat{\Pi}_2^u$
follow from standard arguments. Notice that $E^s(q)\oplus E^c(q)$ and 
$E^u(q)\oplus E^c(q)$ along Loop-0 and Loop-1 (except the small portion 
$\z$), are invariant bundles under $DF^t$. When $\dl$ is small enough,
the inequalities for $\Pi_\ell^s$ ($\ell =1,2$) and $\Pi_\ell^u$ ($\ell =1,3$)
follow. $\Box$
\begin{lemma}\label{acts} $G:\Gamma _{\epsilon ,\ga }\mapsto 
\Gamma_{\epsilon , \ga }$\end{lemma}

Proof: First we show that $G:\Gamma_{\epsilon , \ga }\mapsto 
\Gamma_{\epsilon}$\ .
\[
\| x^s_1(\tau)\| 
=\|g^s(\tau^-,x^u(\tau^-),x^s(\tau^-))\| .
\]
We use the Taylor formula
(\ref{texp}) in some ${\cal B}_j$,
\begin{eqnarray}
& & g^s(\tau^-,x^u(\tau^-),x^s(\tau^-))=
g^s(\tau^-,0,0)+D_2g^s(\tau^-,0,0)x^u(\tau^-) 
+ D_3g^s(\tau^-,0,0)x^s(\tau^-) \nonumber \\ 
& & \quad +\left( \int^1_0\frac{(1-\alpha)}{1!}D^2g^s(\tau^-,\alpha 
x^u(\tau^-),\alpha x^s(\tau^-))d\alpha \right)
(x^u(\tau^-),x^s(\tau^-))^{(2)}.
\label{act1}
\end{eqnarray}
From (\ref{tube4}), we have
\begin{equation}
\label{act2}\| 
\int^1_0\frac{(1-\alpha)}{1!}D^2g^s(\tau^-,\alpha 
x^u(\tau^-),\alpha x^s(\tau^-))d\alpha \| \leq
\frac{1}{2}
\Lambda_2.\end{equation} From Lemma~\ref{Lian},
\begin{equation}\label{act3} \| D_2g^s(\tau^-,0,0)\| <\mu, \quad \| 
D_3g^s(\tau^-,0,0)\|<\frac{1}{2}.\end{equation} For each fixed $T$,
if $\delta$ is sufficiently small, then
\begin{equation}\label{act4} g^s(\tau^-,0,0)\sim {\cal O} 
(\delta),\quad \forall \tau^-\in R.\end{equation} Thus, by
(\ref{act1}-(\ref{act4}), if $\epsilon$ is small enough and, for 
each $\epsilon $, $\delta$ is sufficiently small, then
\begin{equation}\label{act5} \| x^s_1(\tau)\| \leq 
\frac{9}{10}\epsilon ,\quad \forall \tau \in R.\end{equation}
Next we estimate $\|
x^u_1(\tau)\|$. We start with considering 
$g^u(\tau,x^u,x^s(\tau))$ where $\| x^u\| \leq \epsilon$. We 
use the Taylor formula
(\ref{texp}) in some ${\cal B}_j$,
\begin{equation}
g^u(\tau ,x^u,x^s(\tau )) = g^u(\tau ,0,0)+D_2g^u(\tau 
,0,0)x^u +D_3g^u(\tau ,0,0)x^s(\tau ) 
+{\cal O}(\epsilon^2).
\label{act6}
\end{equation}
From 
Lemma~\ref{Lian},
\begin{equation}\label{act7} \| D_3g^u(\tau ,0,0)\| <\mu ,\quad \| \{ 
D_2g^u(\tau,0,0)\}^{-1}\| <\frac{1}{2}.\end{equation} For each
fixed $T$, if $\delta$ is sufficiently small, then
\begin{equation}\label{act8} g^u(\tau ,0,0)\sim 
{\cal O}(\delta),\quad \forall \tau \in R.\end{equation} From 
(\ref{act6}),
\[
x^u=\{ D_2g^u(\tau ,0,0)\}^{-1}\{ 
g^u(\tau , x^u,x^s(\tau ))-g^u(\tau ,0,0) - D_3
g^u(\tau ,0,0)x^s(\tau 
)+{\cal O}(\epsilon^2)\}.
\]
Then by (\ref{act7}) and (\ref{act8}), if $\epsilon$ is
small enough and for each $\epsilon $, $\delta $ is sufficiently small,
\[
\| x^u\|\leq \frac{1}{2}\left(\| g^u(\tau 
,x^u,x^s(\tau ))\| +\frac{1}{2} \epsilon 
\right).
\]
If we take $x^u=x^u_1(\tau )$, we have
\begin{equation}\label{act9} \| x^u_1(\tau )\| \leq 
\frac{3}{4}\epsilon .\end{equation}
Next we show that $G:\Gamma_{\epsilon
,\gamma}\mapsto \Gamma_{\epsilon ,\gamma}$.
\[
(G\sigma )(\tau _\ell) = (\tau 
_\ell ,x^u_1(\tau _\ell ),x^s_1(\tau _\ell)), \quad (\ell =1,2),
\]
where
\begin{eqnarray*}
& &\tau _\ell =f(\tau^-_\ell ,x^u(\tau^-_\ell ),x^s(\tau^-_\ell)),\\ 
& & x^s_1(\tau_\ell)=g^s(\tau^-_\ell ,x^u(\tau^-_\ell),x^s(\tau^-_\ell)),\\ 
& & f(\tau _\ell ,x^u_1(\tau_\ell),x^s(\tau_\ell ))=\tau^+_\ell,\\ 
& & g^u(\tau_\ell ,x^u_1(\tau_\ell),x^s(\tau_\ell))=x^u(\tau^+_\ell). 
\end{eqnarray*}
We have
\begin{eqnarray*}
\| x^s_1(\tau_1) -x^s_1(\tau_2)\| &=& \| g^s(\tau^-_1,x^u(\tau^-_1),x^s(\tau^-_1))-g^s(\tau^-_2,x^u(\tau^-_2),x^s(\tau^-_2))\|\\ 
& \leq & \| g^s(\tau^-_1,x^u(\tau^-_1),x^s(\tau^-_1))-g^s(\tau^-_2,x^u(\tau^-_1),x^s(\tau^-_1))\|\\ 
& &+\| g^s(\tau^-_2,x^u(\tau^-_1),x^s(\tau^-_1))-g^s(\tau_2^-,x^u(\tau^-_2),x^s(\tau^-_1))\|\\ 
& &+\| g^s(\tau^-_2,x^u(\tau^-_2),x^s(\tau^-_1))-g^s(\tau^-_2,x^u(\tau^-_2),x^s(\tau^-_2))\| \\ 
&\leq & \Pi^s_1|\tau^-_1-\tau^-_2|+\Pi_2^s\gamma 
|\tau^-_1-\tau^-_2|+\Pi^s_3\gamma |\tau^-_1-\tau^-_2|,
\end{eqnarray*}
by (\ref{MVI}). That is,
\begin{equation}\label{act10} \| x^s_1(\tau_1)-x^s_1(\tau_2)\| 
\leq (\Pi_1^s+\Pi^s_2\gamma + \Pi_3^s\gamma
)|\tau^-_1-\tau^-_2|.\end{equation}
Next we need to estimate $|\tau^-_1-\tau^-_2|$ in terms of $|\tau _1-\tau_2|$.
\begin{eqnarray}
|\tau_1-\tau_2| &=& |f(\tau^-_1,x^u(\tau^-_1),x^s(\tau^-_1))-f(\tau^-_2,x^u(\tau^-_2),x^s(\tau^-_2))| \nonumber \\ 
&\geq & |f(\tau^-_1,x^u(\tau^-_1),x^s(\tau_1^-))-f(\tau^-_2,x^u(\tau^-_1),x^s(\tau^-_1))| \nonumber \\
& & -|f(\tau^-_2,x^u(\tau^-_1),x^s(\tau^-_1))-f(\tau^-_2,x^u(\tau 
^-_2),x^s(\tau^-_1))|\nonumber  \\
& & -|f(\tau^-_2,x^u(\tau^-_2),x^s(\tau^-_1))-f(\tau^-_2,x^u(\tau^-_2),x^s(\tau^-_2))| \nonumber \\ 
&\geq & 
|f(\tau^-_1,x^u(\tau^-_1),x^s(\tau^-_1))-f(\tau^-_2,x^u(\tau^-_1),x^s(\tau^-_1))|-2\Lambda_1\gamma |\tau^-_1-\tau^-_2|, 
\label{act11}
\end{eqnarray}
where $\Lambda_1$ is defined in (\ref{tube4}). We consider the map
\[
\varphi(\tau^-)=f(\tau^-,x^u(\tau^-_1),x^s(\tau^-_1)),\quad 
\tau^-_1\ \mbox{fixed},
\] 
which is 1-to-1. Then
\[
\tau^-=\varphi^{-1}(f(\tau^-,x^u(\tau^-_1),x^s(\tau^-_1))).
\]
Thus
\[
D\varphi^{-1}(f(\tau^-,x^u(\tau^-_1),x^s(\tau^-_1)))=\{D_1f(\tau^-,x^u(\tau^-_1),x^s(\tau^-_1))\}^{-1}.
\]
By (\ref{MVI}),
\begin{equation}\label{act12} |\tau^-_1-\tau^-_2|\leq
\Lambda_1|f(\tau^-_1,x^u(\tau^-_1),x^s(\tau^-_1))-f(\tau^-_2,x^u(\tau^-_1),x^s(\tau^-_1))|.\end{equation} 
Then we have from
(\ref{act11}) and (\ref{act12}),
\begin{equation}\label{act13} |\tau_1-\tau_2|\geq 
(\Lambda^{-1}_1-2\Lambda_1\gamma )|\tau^-_1-\tau_2^-|.\end{equation} 
Thus from
(\ref{act10}) and (\ref{act13}),
\begin{eqnarray}
\| x^s_1(\tau_1)-x^s_1(\tau_2)\| &\leq& ( \Pi^s_1+\Pi^s_2\gamma 
+\Pi^s_3\gamma
)(\Lambda^{-1}_1-2\Lambda_1\gamma)^{-1}|\tau_1-\tau_2| \nonumber \\ 
&=& \frac{\Lambda_1\Pi^s_1+\Lambda_1\Pi^s_2\gamma
+\Lambda_1\Pi^s_3\gamma}{1-2\Lambda^2_1\gamma}|\tau_1-\tau_2|.
\label{act14}
\end{eqnarray}
Then by Lemma~\ref{Lian},
\begin{equation}\label{act15} \| x^s_1(\tau_1)-x^s_1(\tau_2)\| 
\leq \frac{\Lambda_1\mu(1+\gamma
)+\frac{1}{2}\gamma}{1-2\Lambda^2_1\gamma}|\tau_1-\tau_2|.\end{equation} 
First we choose $\gamma$ small enough such that
$1-2\Lambda^2_1\gamma >\frac{3}{4}$. Then for each $\gamma$ we choose 
$\mu$ small enough such that
\[
\mu <\frac{7}{40}\frac{\gamma }{\Lambda_1(1+\gamma )}.
\]
With these choices, we have
\begin{equation}\label{act16} \| x^s_1(\tau_1)-x^s_1(\tau_2)\| 
<\frac{9}{10}\gamma |\tau_1-\tau _2|.\end{equation}
Next we
estimate $\| x^u_1(\tau_1)-x^u_1(\tau_2)\|$.
\begin{eqnarray}
\| x^u(\tau^+_1)-x^u(\tau^+_2)\| &=& \|
g^u(\tau_1,x^u_1(\tau_1),x^s(\tau_1))-g^u(\tau_2,x^u_1(\tau_2),x^s(\tau_2))\|
\nonumber \\ 
&\geq & \| g^u(\tau_1,x^u_1(\tau_1),x^s(\tau_1))-g^u(\tau_1,
x^u_1(\tau_2),x^s(\tau_1))\| \nonumber\\ 
& & -\| g^u(\tau_1,x^u_1(\tau_2),x^s(\tau_1))-g^u(\tau_2,x^u_1(\tau_2),x^s(\tau_1))\|\nonumber \\ 
& & -\| g^u(\tau_2,x^u_1(\tau_2),x^s(\tau_1))-g^u(\tau_2,x^u_1(\tau_2),x^s(\tau_2))\| \nonumber \\ 
&\geq & \| g^u(\tau_1,x^u_1(\tau_1),x^s(\tau_1))-g^u(\tau_1,x^u_1(\tau_2),x^s(\tau_1))\| \nonumber \\ 
& & -\Pi^u_1|\tau_1-\tau_2|-\Pi^u_3\gamma |\tau_1-\tau_2|.
\label{act17}
\end{eqnarray}
Notice also that
\begin{equation}
\| g^u(\tau_1,x^u_1(\tau_1),x^s(\tau_1))-g^u(\tau, 
x^u_1(\tau_2),x^s(\tau_1))\| \geq
( \widehat{\Pi}^u_2)^{-1}\| 
x^u_1(\tau_1)-x^u_1(\tau_2)\|.
\label{act18}
\end{equation}
Thus from (\ref{act17}) and (\ref{act18}), one has 
\begin{eqnarray}
x^u_1(\tau_1)-x^u_1(\tau_2)\| &\leq & \widehat{\Pi}^u_2 \bigg [\|
x^u(\tau^+_1)-x^u(\tau^+_2)\| + (\Pi^u_1+\Pi^u_3\gamma 
)|\tau_1-\tau_2| \bigg ] \nonumber \\ 
&\leq & \widehat{\Pi}^u_2\bigg [\gamma \
\tau^+_1-\tau^+_2|+(\Pi^u_1+\Pi^u_3\gamma 
)|\tau_1-\tau_2| \bigg ].
\label{act19}
\end{eqnarray}
Next we estimate $|\tau^+_1-\tau^+_2|$ in terms
of $|\tau_1-\tau_2|$.
\begin{eqnarray}
|\tau^+_1-\tau^+_2| &=& |f(\tau_1,x^u_1(\tau_1),x^s(\tau_1))-f(\tau_2,x^u_1(\tau_2),x^s(\tau_2))| \nonumber \\ 
&\leq & \Lambda_1(|\tau_1-\tau_2|+\| x^u_1(\tau_1)-x^u_1(\tau_2)\| 
+\gamma |\tau_1-\tau_2|).
\label{act20}
\end{eqnarray}
From (\ref{act19}) and (\ref{act20}),
\begin{equation}
\| x^u_1(\tau_1)-x^u_1(\tau_2)\| \leq 
(1-\Lambda_1\widehat{\Pi}_2^u\gamma
)^{-1}\bigg [\Lambda_1\widehat{\Pi}^u_2(1+\gamma )\gamma
+\widehat{\Pi}^u_2(\Pi^u_1+\Pi^u_3\gamma)\bigg ]|\tau_1-\tau_2|.
\label{act21}
\end{equation}
By Lemma~\ref{Lian},
\begin{equation}\label{act22} \| x^u_1(\tau_1)-x^u_1(\tau_2)\| 
\leq \left( 1-\frac{1}{2}\gamma \right)^{-1}\left[
\frac{1}{2}(1+\gamma )\gamma+\frac{1}{2}\mu (1+\gamma)\right] 
|\tau_1-\tau_2|.\end{equation}
First we choose $\gamma$ small enough such that 
$1-\frac{1}{2}\gamma >\frac{15}{16}$. Then for each $\gamma$ we 
choose $\mu $ small
enough such that
\[
\mu <\frac{3}{16} (1+\gamma)^{-1}\gamma.
\]
With these choices, we have
\begin{equation}\label{act23} \| x^u_1(\tau_1)-x^u_1(\tau_2)\| 
<\frac{9}{10}\gamma |\tau_1-\tau_2|.\end{equation}
The proof of the lemma is completed. $\Box$

\begin{lemma}\label{contr} $G$ is a contraction on $\Gamma_{\epsilon 
,\gamma}$ in $C^0$ norm.\end{lemma}

Proof: Let $\sigma^{(\ell)}(\ell=1,2)$ be any two sections in 
$\Gamma_{\epsilon ,\gamma}$, and let
\[
\sigma ^{(\ell)}(\tau) =(\tau, 
x^{(u,\ell)}(\tau),x^{(s,\ell)}(\tau)),\ \  (\ell=1,2).
\]
Let
\[
(G\sigma^{(\ell)})(\tau )=(\tau, 
x_1^{(u,\ell)}(\tau ),x_1^{(s,\ell)}(\tau )),\ \ (\ell =1,2),
\]
where
\begin{eqnarray*}
& & \tau =f(\tau^{(-,\ell)},x^{(u,\ell)}(\tau^{(-,\ell)}),x^{(s,\ell)}(\tau^{(-,\ell)})), \ (\ell=1,2),\\ 
& & x^{(s,\ell)}_1(\tau)=g^s(\tau^{(-,\ell)},x^{(u,\ell)}(\tau^{(-,\ell)}),x^{(s,\ell)}(\tau^{(-,\ell)})),\ (\ell=1,2),\\
& & f(\tau,x^{(u,\ell)}_1(\tau),x^{(s,\ell)}(\tau ))=\tau 
^{(+,\ell)}, \ (\ell=1,2)\\ 
& & g^u(\tau ,x^{(u,\ell)}_1(\tau),x^{(s,\ell)}(\tau))=x^{(u,\ell)}(\tau ^{(+,\ell)}),\ (\ell=1,2).
\end{eqnarray*}
First we estimate $\| x^{(s,1)}_1(\tau )-x ^{(s,2)}_1(\tau )\|$.
\begin{eqnarray}
& & \quad \| x_1^{(s,1)}(\tau)-x^{(s,2)}_1(\tau) \| \nonumber \\
&=& \| g^s(\tau^{(-,1)},x^{(u,1)}(\tau^{(-,1)}),x^{(s,1)}(\tau^{(-,1)})) 
-g^s(\tau^{(-,2)},x^{(u,2)}(\tau^{(-,2)}),x^{(s,2)}(\tau^{(-,2)}))\|
\nonumber\\ 
& \leq &\| g^s(\tau^{(-,1)},x^{(u,1)}(\tau^{(-,1)}),x^{(s,1)}(\tau^{(-,1)}))
-g^s(\tau^{(-,2)},x^{(u,1)}(\tau^{(-,1)}),x^{(s,1)}(\tau^{(-,1)}))\|
\nonumber\\ 
& & +\| g^s(\tau^{(-,2)},x^{(u,1)}(\tau^{(-,1)}),x^{(s,1)} (\tau^{(-,1)}))
-g^s(\tau^{(-,2)},x^{(u,2)}(\tau^{(-,1)}),x^{(s,1)}(\tau^{(-,1)}))\|
\nonumber\\
& & +\|g^s(\tau^{(-,2)}, x^{(u,2)} (\tau^{(-,1)}), x^{(s,1)} 
(\tau^{(-,1)}))
-g^s(\tau^{(-,2}, x^{(u,2)}(\tau^{(-,2)}),x^{(s,1)}(\tau^{(-,1)}))\| 
\nonumber\\ 
& & +\|g^s(\tau^{(-,2)},x^{(u,2)}(\tau^{(-,2)}),x^{(s,1)}(\tau^{(-,1)}))
-g^s(\tau^{(-,2)},x^{(u,2)}(\tau^{(-,2)}),x^{(s,2)}(\tau^{(-,1)}))\|
\nonumber\\ 
& & +\|g^s(\tau^{(-,2)},x^{(u,2)}(\tau^{(-,2)}),x^{(s,2)}(\tau^{(-,1)})) 
-g^s(\tau^{(-,2)},x^{(u,2)}(\tau^{(-,2)}), x^{(s,2)}(\tau^{(-,2)}))\| 
\nonumber\\ 
&\leq & \Pi^s_1 |\tau^{(-,1)}-\tau^{(-,2)}|+\Pi^s_2\| 
x^{(u,1)}(\tau^{(-,1)})-x^{(u,2)}(\tau ^{(-,1)})\|
+ \Pi^s_2 \ga |\tau^{(-,1)}-\tau^{(-,2)}|\nonumber\\ 
& & + \Pi^s_3\| x^{(s,1)}(\tau^{(-,1)})-x^{(s,2)}(\tau ^{(-,1)})\|
+\Pi^s_3 \ga |\tau^{(-,1)}-\tau^{(-,2)}|\nonumber\\
& = &(\Pi^s_1+\Pi^s_2 \gamma +\Pi^s_3\gamma)|\tau^{(-,1)}-\tau^{(-,2)}|
+ \Pi^s_2\| x^{(u,1)}(\tau^{(-,1)})-x^{(u,2)}(\tau^{(-,1)})\|
\nonumber\\
& & +\Pi^s_3\| x^{(s,1)}(\tau^{(-,1)})-x^{(s,2)}(\tau^{(-,1)})\|.
\label{ctr1} 
\end{eqnarray}
Next we estimate $|\tau^{(-,1)}-\tau^{(-,2)}|$. Notice that,
\begin{eqnarray*}
& & \| f(\tau^{(-,1)},x^{(u,1)}(\tau^{(-,1)}),x^{(s,1)}(\tau^{(-,1)}))-
f(\tau^{(-,2)},x^{(u,1)}(\tau^{(-,1)}),x^{(s,1)}(\tau^{(-,1)}))\| \\
&=& \| f(\tau^{(-,2)},x^{(u,2)}(\tau^{(-,2)}),x^{(s,2)}(\tau^{(-,2)}))
- f(\tau^{(-,2)},
x^{(u,1)}(\tau^{(-,1)}),x^{(s,1)}(\tau^{(-,1)}))\|,
\end{eqnarray*}
where
\begin{eqnarray*}
& & \| f(\tau^{(-,1)},x^{(u,1)}(\tau^{(-,1)}),x^{(s,1)}(\tau^{(-,1)}))-f(\tau^{(-,2)},x^{(u,1)}(\tau^{(-,1)}),x^{(s,1)}(\tau^{(-,1)}))\| \\
& & \quad \geq \Lambda^{-1}_1|\tau^{(-,1)}-\tau^{(-,2)}|,
\end{eqnarray*}
and
\begin{eqnarray*}
& & \| f(\tau^{(-,2)},x^{(u,2)}(\tau^{(-,2)}),x^{(s,2)}
(\tau^{(-,2)}))-f(\tau^{(-,2)},x^{(u,1)}(\tau^{(-,1)}),x^{(s,1)}
(\tau^{(-,1)}))\| \\
& & \quad \leq  \Lambda_1 \bigg ( 2\gamma |\tau^{(-,1)}-\tau^{(-,2)}|+\| 
x^{(u,1)}(\tau^{(-,1)})-x^{(u,2)}(\tau^{(-,1)})\| \\
& & \quad + \| x^{(s,1)}(\tau^{(-,1)})-x^{(s,2)}(\tau 
^{(-,1)})\| \bigg ) .
\end{eqnarray*}
Then
\begin{eqnarray}
|\tau^{(-,1)}-\tau^{(-,2)}|&\leq &
\frac{\Lambda_1}{\Lambda^{-1}_1-2\gamma \Lambda_1} \bigg [\|
x^{(u,1)}(\tau^{(-,1)})-x^{(u,2)}(\tau^{(-,1)})\| \nonumber \\
&+& \| x^{(s,1)} (\tau^{(-,1)})- x^{(s,2)} (\tau^{(-,1)})\| \bigg ].
\label{ctr2} 
\end{eqnarray}
From (\ref{ctr1}) and (\ref{ctr2}),
\[
\|x^{(s,1)}_1(\tau 
)-x^{(s,2)}_1(\tau )\| \leq c_s\|
x^{(s,1)}(\tau^{(-,1)})-x^{(s,2)}(\tau^{(-,1)})\| +c_u\| 
x^{(u,1)}(\tau^{(-,1)})-x^{(u,2)}(\tau^{(-,1)})\|,
\]
where
\[
c_s=\frac{1}{2} +\frac{\mu (1+\gamma 
)\Lambda^2_1+\frac{1}{2}\gamma \Lambda_1}{1-2\gamma \Lambda^2_1}, 
\quad c_u=\mu
+\frac{\mu (1+\gamma )\Lambda^2_1+\frac{1}{2}\gamma 
\Lambda_1}{1-2\gamma \Lambda^2_1}.
\]
If $\gamma$ and $\mu$ are small enough, then
\begin{equation}\label{ctr3} \| 
x^{(s,1)}_1(\tau)-x_1^{(s,2)}(\tau)\| \leq \frac{9}{10}\| 
\sigma^{(1)}-\sigma^{(2)}\|
_0.
\end{equation}
Next we estimate $\| x^{(u,1)}_1(\tau )-x_1^{(u,2)}(\tau )\|$.
\begin{eqnarray*}
& & \quad \| x^{(u,1)}(\tau^{(+,1)})-x^{(u,2)}(\tau^{(+,2)})\| \\
&=& \| g^u(\tau, x_1^{(u,1)}(\tau),x^{(s,1)}(\tau ))
 - g^u(\tau ,x_1^{(u,2)}(\tau ), x^{(s,2)}(\tau ))\| \\
&\geq & \| g^u(\tau, x_1^{(u,1)}(\tau ),x^{(s,1)}(\tau 
))-g^u(\tau ,x_1^{(u,2)}(\tau ) ,x^{(s,1)}(\tau))\|\\
& & -\|g^u(\tau ,x^{(u,2)}_1(\tau),x^{(s,1)}(\tau))-g^u(\tau 
,x_1^{(u,2)}(\tau),x^{(s,2)}(\tau ))\|\\
&\geq & (\widehat{\Pi}_2^u)^{-1}\| x^{(u, 1)}_1(\tau 
)-x^{(u,2)}_1(\tau )\| -\Pi^u_3\| 
x^{(s,1)}(\tau)-x^{(s,2)}(\tau)\|.
\end{eqnarray*}
Then
\begin{eqnarray}
& & \quad \| x^{(u,1)}_1(\tau )-x^{(u,2)}_1(\tau )\| \\
&\leq &\widehat{\Pi}_2^u \bigg [\|
x^{(u,1)}(\tau^{(+,1)})-x^{(u,2)}(\tau^{(+,2)})\| 
+\Pi^u_3\| x^{(s,1)}(\tau)-x^{(s,2)}(\tau )\| \bigg ]\nonumber \\
&\leq &\widehat{\Pi}^u_2 \bigg [\| 
x^{(u,1)}(\tau^{(+,1)})-x^{(u,2)}(\tau^{(+,1)})\| +\gamma 
|\tau^{(+,1)}-\tau^{(+,2)}|\nonumber \\
& & + \Pi^u_3\| x^{(s,1)}(\tau )-x^{(s,2)}(\tau 
)\| \bigg ]. \label{ctr4} 
\end{eqnarray}
Next we estimate $|\tau^{(+,1)}-\tau^{(+,2)}|$.
\begin{eqnarray}
|\tau^{(+,1)}-\tau^{(+,2)}| &=& |f(\tau,x_1^{(u,1)}(\tau),x^{(s,1)}(\tau))
-f(\tau,x^{(u,2)}_1(\tau),x^{(s,2)}(\tau))| \nonumber \\
&\leq &\Lambda_1 \bigg [\| x^{(u,1)}_1(\tau)-x^{(u,2)}_1(\tau)\| +\| 
x^{(s,1)}(\tau)-x^{(s,2)}(\tau)\| \bigg ]. \label{ctr5}
\end{eqnarray}
From (\ref{ctr4}) and (\ref{ctr5}),
\begin{eqnarray*}
\| x^{(u,1)}_1(\tau)-x^{(u,2)}_1(\tau)\| &\leq &
\frac{\widehat{\Pi}^u_2}{1-\gamma
\Lambda_1\widehat{\Pi}_2^u}\bigg [\| 
x^{(u,1)}(\tau^{(+,1)})-x^{(u,2)}(\tau^{(+,1)})\|\\
& & + (\Pi^u_3+\gamma \Lambda_1)\| x^{(s,1)}(\tau)-x^{(s,2)}(\tau )\|\bigg ]\\
&\leq & \frac{\frac{1}{2}}{1-\frac{1}{2}\gamma}\bigg [\| 
x^{(u,1)}(\tau^{(+,1)})-x^{(u,2)}(\tau^{(+,1)})\| \\
& & +(\mu+\gamma \Lambda_1)\| x^{(s,1)}(\tau)-x^{(s,2)}(\tau)\| \bigg ].
\end{eqnarray*}
If $\gamma $ and $\mu$ are small enough, then
\begin{equation}\label{ctr6} \| x^{(u,1)}_1(\tau)-x^{(u,2)}_1(\tau)\| 
\leq \frac{9}{10}\| \sigma^{(1)}-\sigma^{(2)}\|
_{C^0}.\end{equation}
By (\ref{ctr3}) and (\ref{ctr6}), we have
\begin{equation}\label{ctr7} \| G\sigma^{(1)}-G\sigma^{(2)}\| 
_{C^0}\leq \frac{9}{10}\| 
\sigma^{(1)}-\sigma^{(2)}\|_{C^0}.\end{equation}
The proof of the lemma is completed. $\Box$
\begin{theorem} 
The graph transform $G$ has a unique fixed point 
$\sigma^*$ in $\Gamma_{\epsilon ,\gamma}$. {\em{Graph}} $\sigma^*$ is an
orbit that $\epsilon$-shadows the $\delta$-pseudo-orbit $\eta_a$.
\end{theorem}

Proof: By Lemmas~\ref{compl}, \ref{acts}, and \ref{contr}, $G$ 
has a unique fixed point $\sigma^*$ in $\Gamma_{\epsilon
,\gamma}$. Assume that $\sigma^*$ has the representation
\[
\sigma^*(\tau)=(\tau 
,x^u(\tau),x^s(\tau)),\ \tau \in R.
\]
Then 
\[
(G\sigma^*)(\tau)=(\tau, 
x^u(\tau),x^s(\tau)),\quad \tau \in R,
\]
where
\begin{eqnarray}
& & f(\tau^-,x^u(\tau^-),x^s(\tau^-))=\tau,\label{cac1} \\
& & g^s(\tau^-,x^u(\tau^-),x^s(\tau^-))=x^s(\tau),\label{cac2}\\
& & f(\tau ,x^u(\tau), x^s(\tau ))=\tau^+,\label{cac3}\\
& & g^u(\tau ,x^u(\tau),x^s(\tau))=x^u(\tau^+).\label{cac4}
\end{eqnarray}
Replacing $\tau$ by $\tau^-$ in (\ref{cac3}) and (\ref{cac4}), we have
\begin{eqnarray}
& & f(\tau^-,x^u(\tau^-),x^s(\tau^-))=\tau,\label{cac5} \\
& & g^u(\tau^-,x^u(\tau^-),x^s(\tau^-))=x^u(\tau).\label{cac6} 
\end{eqnarray}
From (\ref{cac1}), (\ref{cac2}), (\ref{cac5}), and (\ref{cac6}), 
$\forall (\tau^-,x^u(\tau^-),x^s(\tau^-))\in \sigma^*$, there
exists a unique $(\tau, x^u(\tau),x^s(\tau))\in \sigma^*$ such that
\[
F^T(\tau^-,x^u(\tau^-),x^s(\tau^-))=(\tau,x^u(\tau),x^s(\tau)),
\]
i.e.
\[
F^T(\mbox{Graph}\ \sigma^*)\subset \mbox{Graph}\ \sigma^*;
\]
$\forall (\tau ,x^u(\tau),x^s(\tau))\in \sigma^*$, there exists 
a unique $(\tau^-,x^u(\tau^-),x^s(\tau^-))\in \sigma^*$
such that
\[
(\tau ,x^u(\tau),x^s(\tau))=F^T(\tau^-,x^u(\tau^-),x^s(\tau^-)),
\]
i.e.
\[
\mbox{Graph} \ \sigma^*\subset F^T(\mbox{Graph}\  \sigma^*).
\]
Thus 
\begin{equation} 
F^T(\mbox{Graph}\  \sigma^*)=\mbox{Graph}\  \sigma^*.
\label{cac7} 
\end{equation}
Let $\sigma$ be a section in $\Gamma_{\epsilon ,\gamma}$ such that
\begin{equation}
F^T(\mbox{Graph}\  \sigma)=\mbox{Graph}\  \sigma.
\label{cac8} 
\end{equation}
We denote $\sigma$ as
\[
\sigma (\tau)=(\tau ,x^u(\tau ), x^s(\tau 
)),\quad \tau \in R.
\]
Then $G\sigma$ is given in (\ref{gt}). By (\ref{cac8}),
\begin{eqnarray}
& & f(\tau^-,x^u(\tau^-),x^s(\tau^-))=\tau,\nonumber \\
& & g^s(\tau^-,x^u(\tau^-),x^s(\tau^-))=x^s_1(\tau)=x^s(\tau),\nonumber \\
& & f(\hat{\tau}^-,x^u_1(\hat{\tau}^-),x^s(\hat{\tau}^-))=\tau 
=f(\tau^-,x^u(\tau^-),x^s(\tau^-)),\label{cac9} \\
& & g^u(\hat{\tau}^-,x^u_1(\hat{\tau}^-),x^s(\hat{\tau}^-))
=x^u(\tau)=g^u(\tau^-,x^u(\tau^-),x^s(\tau^-)).\label{cac10}
\end{eqnarray}
Now we will show that $\hat{\tau}^-=\tau^-$, 
$x^u_1(\tau^-)=x^u(\tau^-)$.
From (\ref{cac10}),
\begin{eqnarray}
& & 0=\| g^u(\hat{\tau}^-,x^u_1
(\hat{\tau}^-),x^s(\hat{\tau}^-))-g^u(\tau^-,x^u(\tau^-),x^s(\tau^-))\|
\nonumber \\
& & \quad \geq  (\widehat{\Pi}^u_2)^{-1}\| 
x^u_1(\tau^-)-x^u(\tau^-)\|-\Lambda_1\gamma 
|\hat{\tau}^--\tau^-| \nonumber \\
& & \quad -\Pi^u_1|\hat{\tau}^--\tau^-|-\Pi_3^u\gamma 
|\hat{\tau}^--\tau^-|. \label{cac11}
\end{eqnarray}
From (\ref{cac9}),
\begin{eqnarray*}
& & 0=|f(\hat{\tau}^-,x^u_1(\hat{\tau}^-),x^s(\hat{\tau}^-))-
f(\tau^-,x^u(\tau^-),x^s(\tau^-))|\\
&\geq &\Lambda^{-1}_1|\hat{\tau}^--\tau^-|-\Lambda_1(\| 
x^u_1(\tau^-)-x^u(\tau^-)\|+2\ga 
|\hat{\tau}^--\tau^-|).
\end{eqnarray*}
Then, 
\begin{equation}
|\hat{\tau}^--\tau^-|\leq 
\frac{\Lambda_1}{\Lambda^{-1}_1-2\gamma \Lambda_1}\|
x^u_1(\tau^-)-x^u(\tau^-)\|.
\label{cac12} 
\end{equation}
Thus, from (\ref{cac11}) and (\ref{cac12}),
\begin{equation}\label{cac13} \left[ 
1-\frac{1}{2}\frac{\Lambda_1\gamma+\mu(1+\gamma)}{\Lambda^{-1}_1-2\gamma 
\Lambda_1}\right] \|
x^u_1(\tau^-)-x^u(\tau^-)\| \leq 0.\end{equation}
When $\gamma$ and $\mu$ are sufficiently small, (\ref{cac13}) implies that
\[
\| x^u_1(\tau^-)-x^u(\tau^-)\|=0,
\]
which in turn implies that, by (\ref{cac12}),
\[
|\hat{\tau}^--\tau^-|=0.
\]
Thus, $\sigma$ is a fixed point of $G$. To summarize, we have 
$\sigma$ is a fixed point of $G$, i.e. $G\sigma=\sigma$, if and only 
if equation (\ref{cac8}) holds. For $t\in [-t_0,t_0]$, $t_0>0$, we 
define $F^t(\mbox{Graph}\  \sigma^*)$ to be the graph of a certain section
$\sigma^*_t$. Since $G\sigma^*=\sigma^*$, by (\ref{act5}), 
(\ref{act9}), (\ref{act16}), and (\ref{act23}),
\[
\| \sigma^*\|_{C^0}\leq \frac{9}{10}\epsilon ,\quad 
\mbox{Lip}\ \{ \sigma^*\}\leq \frac{9}{10} \gamma.
\]
Thus there exists a small $t_0>0$, such that
\[
\| \sigma^*_t\| _{C^0}\leq \epsilon ,\quad \mbox{Lip}\ \{ 
\sigma^*_t\}\leq \gamma ,\quad \forall t\in
[-t_0,t_0],
\]
i.e. $\sigma^*_t\in \Gamma_{\epsilon ,\gamma}$. From (\ref{cac7}),
\begin{equation}\label{cac14} 
F^t(\mbox{Graph} \ \sigma^* )=F^tF^T(\mbox{Graph}\  
\sigma^*)=F^TF^t(\mbox{Graph}\  \sigma^*).\end{equation}
(\ref{cac14}) is equivalent to $G\sigma^*_t=\sigma^*_t$. Then by the 
uniqueness of the fixed point of $G$ in $\Gamma_{\epsilon
,\gamma}$,
\[
\sigma^*_t=\sigma^*,\quad \forall t\in [-t_0,t_0].
\]
Thus,
\begin{equation}\label{cac15} F^t(\mbox{Graph}\  \sigma^*)=\mbox{Graph}\   
\sigma^*,\quad \forall t\in [-t_0,t_0].\end{equation}
Iteration of (\ref{cac15}) leads to
\[
F^t(\mbox{Graph}\  \sigma^*)=\mbox{Graph} \ \sigma^*,\quad \forall 
t\in (-\infty ,\infty).
\]
That is, Graph $\sigma^*$ is an orbit that $\epsilon$-shadows the 
$\delta$-pseudo-orbit $\eta_a$. The proof of the theorem is completed.
$\Box$
\begin{remark}
As curves, the shadowing orbits are Lipschitz, and can be $C^k$ 
smooth for some $k >0$. But this does not mean that the shadowing orbits 
are Lipschitz in time.
\end{remark}

\eqnsection{Chaos}

First we will define a return map $P$. We will use notations from 
Section \ref{PO}. Pick a point $p_*$ on $S$, which is ${\cal O}(1)$ 
away from $p_c$ in $\dl$. At $p_*$, we set up a transversal section 
$\Xi$ to $S$. For any pseudo-orbit $\eta_a$, denote by $h_{a_0}$ 
the portion of the shadowing orbit, that shadows the portion Loop-$a_0$ 
of the pseudo-orbit. Let $q_a$ be the first intersection point of 
$h_{a_0}$ with $\Xi$. Let $\La$ be the set consisting of $q_a$ for all 
doubly infinite sequences $a \in \Sg$. We define the return map $P: 
\La \mapsto \La$ as follows: For any $q_a \in \La$, $P(q_a) = q_{\chi(a)}$.
\begin{theorem}[Chaos Theorem]
The subset $\La \subset \Sg$ is invariant under the return map $P$. 
The action of $P$ on $\La$ is topologically conjugate to the action of 
the shift automorphism $\chi$ on $\Sg$. That is, there exists a 
homeomorphism $\phi: \Sg \mapsto \La$ such that the following diagram 
commutes:
\[
\begin{array}{ccc}
\Sg &\maprightu{\phi} & \Lambda\\
\mapdownl{\chi} & & \mapdownr{P}\\
\Sg & \maprightd{\phi} & \Lambda
\end{array} 
\]
\label{autoth}
\end{theorem}

Proof: The invariance of $\La$ under $P$ follows from the definitions of 
$\La$ and $P$. We define $\phi: \Sg \mapsto \La$ as follows: For any $a \in 
\Sg$, $\phi(a)=q_a$. It is straightforward to show that $\phi$ is a 
homeomorphism, and $P$ and $\chi$ are topologically conjugate. $\Box$

\eqnsection{An Example: A Derivative Nonlinear Schr\"odinger Equation}

Consider the derivative nonlinear Schr\"odinger equation,
\begin{equation}
i q_t = q_{xx} + 2 |q|^2 q +i \e \bigg [ (\frac{9}{16}-|q|^2 )q +\mu 
|\hat{\pa}_x q|^2 \bar{q} \bigg ]\ , \label{derNLS}
\end{equation}
where $q$ is a complex-valued function of two real variables $t$ and $x$,
$\e > 0$ is the perturbation parameter, $\mu$ is a real constant, and
$\hat{\pa}_x $ is a bounded Fourier multiplier,
\[
\hat{\pa}_x q = -\sum_{k=1}^K k \tq_k \sin kx\ , \quad 
\mbox{for} \ q = \sum_{k=0}^\infty \tq_k \cos kx\ ,
\]
and some fixed $K$ (cf: \cite{LMSW96}). Periodic boundary condition 
and even constraint are imposed,
\[
q(t,x+2\pi ) = q(t,x)\ , \ \ q(t,-x)=q(t,x) \ . 
\] 
\begin{theorem}[Transversal Homoclinic Orbit Theorem]
There exists a $\e_0 > 0$, such that 
for any $\e \in (0, \e_0)$, and $|\mu | > 5.8$,
there exist two transversal homoclinic orbits asymptotic to 
the limit cycle $q_c = \frac{3}{4} \exp \{ -i [ \frac{9}{8} t + \ga ]\}$.
\end{theorem}

Proof: Denote by $q_c$ the limit cycle,
\[
q_c = \frac{3}{4} \exp \{ -i [ \frac{9}{8} t + \ga ]\}\ .
\]
The eigenvalue of this limit cycle is,
\[
\la = -\e \frac{9}{16} \pm \sqrt{k^2(\frac{9}{4}-k^2)+\e^2
\bigg ( \frac{3}{4} \bigg )^4}\ , \quad \mbox{for}\ k = 0, 1, \cdot 
\cdot \cdot \ .
\]
There is only one unstable mode. The same argument as in \cite{LMSW96} 
\cite{Li01a} shows that the size of the stable manifold of the limit cycle is 
${\cal O}(\sqrt{\e})$. Also the same argument as in \cite{LW97} 
\cite{LMSW96} \cite{Li01a} shows that the Fenichel's persistent invaraint 
manifold theorem and fiber theorem are true. As a result, there exist 
codimension 
1 center-stable and center-unstable manifolds, codimension 2 
center manifold, together with stable and unstable fibrations. Thus if the
Melnikov measurement is successful, that is, there exists an orbit 
in the intersection of the unstable manifold of the limit cycle and 
the center-stable manifold, then the orbit will be a homoclinic orbit 
asymptotic to the limit cycle, due to the fact that the size of the 
stable manifold of the limit cycle is ${\cal O}(\sqrt{\e})$. The Melnikov 
function is given as,
\begin{eqnarray*}
M &=& \int_{-\infty}^{\infty}\int_0^{2\pi} \bigg \{ \frac{\dl F_1}{\dl q}
\bigg [ (\frac{9}{16} - |q|^2)q +\mu |\hat{\pa}_x q|^2 \bar{q} \bigg ] \\
&+& \frac{\dl F_1}{\dl \bar{q}}
\bigg [ (\frac{9}{16} - |q|^2)\bar{q} +\mu |\hat{\pa}_x q|^2 q \bigg ]
\bigg \} dx \ d \tau\ ,
\end{eqnarray*}
where $F_1$ is defined in \cite{LMSW96} \cite{Li01a},
\[
\left ( \begin{array}{c} \frac{\dl F_1}{\dl q} \\ 
\frac{\dl F_1}{\dl \bar{q}} \end{array} \right) \sim 
(|u_1|^2+|u_2|^2)^{-2} \left ( \begin{array}{c} \overline{q_c} 
\ \overline{u_1}^{\ 2} \\ 
-q_c \ \overline{u_2}^{\ 2} \end{array} \right )\ ,
\]
and 
\begin{eqnarray*}
u_1 &=& \cosh \frac{\tau}{2} \cos z - i \sinh \frac{\tau}{2} \sin z \ , \\
u_2 &=& -\sinh \frac{\tau}{2} \cos (z-\vth_0) + i \cosh \frac{\tau}{2} 
\sin (z-\vth_0) \ ,
\end{eqnarray*}
and
\begin{eqnarray*}
q &=& q_c \bigg [ 1 + \sin \vth_0 \ \mbox{sech} \tau \cos x \bigg ]^{-1} \\ 
& & \bigg [ \cos 2 \vth_0 - i \sin 2 \vth_0 \tanh \tau -
\sin \vth_0 \ \mbox{sech} \tau \cos x \bigg ]
\end{eqnarray*}
\[
\tau = \frac{\sqrt{5}}{2}t - \rho\ , \quad \vth_0 = \arctan 
\frac{\sqrt{5}}{2}\ ,
\quad z= \frac{x}{2} + \frac{1}{2}(\arctan \frac{\sqrt{5}}{2} - 
\frac{\pi}{2})\ ,
\]
where $\rho$ is the B\"acklund parameter. $q_c$ can be rewritten as
\[
q_c=\frac{3}{4} \exp \{-i[\frac{9}{4\sqrt{5}}\tau -\tilde{\ga} ]\}\ ,
\]
where
\[
\tilde{\ga} =-(\ga + \frac{9}{4\sqrt{5}}\rho )\ .
\]
The solutions for $M=0$ are given by,
\[
\cos 2 \tilde{\ga} = \frac{5.8}{\mu }\ .
\]
This completes the proof. $\Box$

\begin{theorem}[Chaos Theorem]
There exists a $\e_0 > 0$, such that 
for any $\e \in (0, \e_0)$, and $|\mu | > 5.8$,
Theorem \ref{autoth} holds for the derivative nonlinear 
Schr\"odinger equation (\ref{derNLS}).
\end{theorem}

Proof: Arguments as in \cite{Li01a} show that the transversal 
homoclinic orbit is a classical solution. Thus, Assumption (A1) 
is valid. Assumption (A2) follows from the standard arguments 
in \cite{LW97} \cite{LMSW96} \cite{Li01a}. Since the perturbation 
in (\ref{derNLS}) is bounded, Assumption (A3) follows from standard 
arguments. $\Box$

\eqnsection{Appendix: Chaos in Non-Autonomous Perturbed Soliton Equations}

Consider the periodically perturbed sine-Gordon equation,

\begin{equation}
u_{tt}=c^2 u_{xx}+\sin u+\epsilon [-a u_t+u^3 \chi(\| u\|)\cos t],
\label{PSG}
\end{equation} 
where
\[
\chi(\| u\|)=\left\{ \begin{array}{ll} 1, & \| 
u\| \leq M,\\ 0, & \| u\| \geq 2M,\end{array}\right.
\]
for $M<\| u\| <2M$, $\chi (\| u\|)$ is a smooth bump function (see Figure
\ref{mfu}),
under odd periodic boundary condition,
\[
u(x+2\pi ,t)=u(x,t),\quad 
u(x,t)=-u(x,t),
\]
$\frac{1}{4}<c^2<1$, $a>0$, $\epsilon$ is a small 
perturbation parameter.
\begin{figure}
\vspace{1.5in}
\caption{The bump function.}
\label{mfu}
\end{figure}
\begin{theorem}[\cite{LMSW96}, \cite{SZ00}] There exists an 
interval $I\subset R^{+}$ such that for any $a\in I$, there exists a
transversal homoclinic orbit
$u=\xi (x,t)$ asymptotic to $0$ in $H^{1}$.
\end{theorem}
Denote by $P$ the time-$2\pi$ Poincar\'e map of the system 
(\ref{PSG}).  Then $P$ is a $C^{2}$-diffeomorphism on $H^{1}$ 
\cite{LMSW96} (and  references thereof). Under $P$, the transversal 
homoclinic orbit
$u=\xi (x,t)$ changes into the transversal homoclinic orbit $\{
\xi_{j}(x)\}_{j\in Z}$ asymptotic to $0$. Using shadowing 
lemma, Bernoulli shift dynamics can be established in the
neighborhood of the transversal homoclinic orbit. This has been done 
by H. Steinlein and H. O. Walther~\cite{SW89,SW90} and D.
Henry~\cite{Hen94} in infinite dimensions. The theorem stated 
specifically for the perturbed sine-Gordon system (\ref{PSG}) can be
described as follows.

\begin{definition} Denote by $\Sigma_m \ (m\geq 2)$ the set of doubly 
infinite sequences
\[
k=(\ldots , k_{-1}, k_{0}, k_{1}, \ldots )
\]
where $k_{j}\in \{ 1,2, \ldots , m\}$. So $\Sigma_{m}=\{
1,2, \ldots ,  m\}^{Z}$.
\end{definition}

We give the set $\{ 1,2, \ldots ,m\}$ the discrete topology and
$\Sigma_{m}$ the product topology. The Bernoulli shift $\beta 
:\Sigma_{m}\to \Sigma_{m}$ is the homeomorphism defined by
\[
[\beta (k)]_{j}=k_{j+1}.
\]

\begin{theorem} There is an integer $\ell$ and a homeomorphism $\phi$ 
of $\Sigma_{m}$ onto a compact Cantor subset $\Lambda $ of
$H^{1}$.
$\Lambda $ is invariant under the time-$2\pi$ Poincar\'e map $P$ of 
the perturbed sine-Gordon equation (\ref{PSG}). The 
action of $P^{\ell}$ on $\Lambda$ is topologically conjugate to
the  action of $\beta$ on
$\Sigma_{m}:P^{\ell} \circ \phi =\phi \circ
\beta$. That is, the following diagram commutes:
\[
\begin{array}{ccc}
\Sg_m &\maprightu{\phi} & \Lambda\\
\mapdownl{\beta} & & \mapdownr{P}\\
\Sg_m & \maprightd{\phi} & \Lambda
\end{array} 
\]
\end{theorem}

\begin{remark} In finite dimensions, this theorem was proved by K. 
Palmer using shadowing lemmas \cite{Pal84,Pal88}. This is the
well-known Smale horseshoe theorem \cite{Sma65}. See also related 
works \cite{Zen95,Zen97}. In infinite dimensions, the above theorem
was  proved by H. Steinlein and H.O. Walther~\cite{SW89,SW90} and D. 
Henry~\cite{Hen94}. See also related works
\cite{HL86,CLP89,CLP89a,Bla86}.\end{remark}